\documentclass[english,conference, two column]{IEEEtran}
\pdfoutput=1
\usepackage[T1]{fontenc}
\usepackage[latin9]{inputenc}
\usepackage{verbatim}
\usepackage{units}
\usepackage{amsthm}
\usepackage{amssymb}
\usepackage{graphicx}
\usepackage{esint}

\makeatletter
\theoremstyle{definition}
\newtheorem{assumption}{Assumption}
  \theoremstyle{plain}
  \newtheorem{lem}{\protect\lemmaname}
  \theoremstyle{plain}
  \newtheorem{thm}{\protect\theoremname}
  \theoremstyle{remark}
  \newtheorem{rem}{\protect\remarkname}

\usepackage{fouridx}
\usepackage{cite}
\IEEEoverridecommandlockouts

\makeatother

\usepackage{babel}
\providecommand{\lemmaname}{Lemma}
\providecommand{\remarkname}{Remark}
\providecommand{\theoremname}{Theorem}

\begin{document}

\title{Online Approximate Optimal Path-Following for a Kinematic Unicycle%
\thanks{Patrick Walters, Rushikesh Kamalapurkar, Lindsey Andrews and Warren
E. Dixon are with the Department of Mechanical and Aerospace Engineering,
University of Florida, Gainesville, FL, USA. Email: \{walters8, rkamalapurkar,
landr010, wdixon\}@ufl{}.edu%
}%
\thanks{This research is supported in part by NSF award numbers 0901491, 1161260,
1217908, ONR grant number N00014-13-1-0151, and a contract with the
AFRL Mathematical Modeling and Optimization Institute. Any opinions,
findings and conclusions or recommendations expressed in this material
are those of the authors and do not necessarily reflect the views
of the sponsoring agency.%
}}

\author{Patrick Walters, Rushikesh Kamalapurkar, Lindsey Andrews, and Warren
E. Dixon}
\maketitle
\begin{abstract}
Online approximation of an infinite horizon optimal path-following
strategy for a kinematic unicycle is considered. The solution to the
optimal control problem is approximated using an approximate dynamic
programming technique that uses concurrent-learning-based adaptive
update laws to estimate the unknown value function. The developed
controller overcomes challenges with the approximation of the infinite
horizon value function using an auxiliary function that describes
the motion of a virtual target on the desired path. The developed
controller guarantees uniformly ultimately bounded (UUB) convergence
of the vehicle to a desired path while maintaining a desired speed
profile and UUB convergence of the approximate policy to the optimal
policy. Simulation results are included to demonstrate the controller's
performance.
\end{abstract}

\section{Introduction}

The goal of a mobile robot feedback controller can be classified into
three categories: point regulation, trajectory tracking, or path following.
Point regulation refers to the stabilization of a dynamical system
about a desired state. Trajectory tracking requires a dynamical system
to track a time parametrized reference trajectory. Path-following
involves convergence of the system state to a given path at a desired
speed profile without temporal constraints. Path-following heuristically
yields smoother convergence to the desired path and reduces the risk
of control saturation \cite{Lapierre.Soetanto.ea2003}. A path-following
control structure can also alleviate difficulties in the control of
nonholonomic vehicles \cite{Morin.Samson2008}. Path-following control
is particularly useful for mobile robots with objectives that emphasize
path convergence and maintaining a desired speed profile (cf. \cite{Lapierre.Soetanto2007,Encarnacao.Pascoal2000,Lapierre2008,Morro.Sgorbissa.ea2011,Dacic.Nesic.ea2007,Aguiar2007,Sarkar.Yun.ea1993}).

To improve path-following performance, optimal control techniques
have been applied to path-following. The result in \cite{Borhaug.Pettersen.ea2006}
combines line-of-sight guidance and model predictive control (MPC)
to optimally follow straight line segments. In \cite{Kanjanawanishkul.Zell2009},
the MPC structure is used to develop a controller for an omnidirectional
robot with dynamics linearized about the desired path. Nonlinear MPC
is used in \cite{Faulwasser.Findeisen2009} to develop an optimal
path-following controller for a general mobile robot model over a
finite time-horizon. In \cite{Sharp2006}, the path-following of a
motorcycle is considered using a linear optimal preview control scheme.

Approximate dynamic programming-based (ADP-based) techniques have
been used to approximate optimal control policies for regulation (cf.
\cite{Vamvoudakis2010,Bhasin.Kamalapurkar.ea2013a,Kamalapurkar.Walters.ea2013})
and trajectory tracking (cf. \cite{Dierks.Jagannathan2010,Zhang.Cui.ea2011,arxivKamalapurkar.Dinh.ea2013a}).
ADP stems from the concept of reinforcement learning and Bellman's
principle of optimality where the solution to the Hamilton-Jacobi-Bellman
(HJB) equation is approximated using parametric function approximation
techniques, and an actor-critic structure is used to estimate the
unknown parameters. Various methods have been proposed in \cite{Vamvoudakis2010,Bhasin.Kamalapurkar.ea2013a,Kamalapurkar.Walters.ea2013,Dierks.Jagannathan2010,Zhang.Cui.ea2011,arxivKamalapurkar.Dinh.ea2013a,Beard1997,Abu-Khalaf2002,Vrabie2009,Vamvoudakis2009}
to approximate the solution to the HJB equation. For an infinite horizon
regulation problem, function approximation techniques, such as neural
networks (NNs), are used to approximate the value function and the
optimal policy.

Motivated by the desire to develop a nonlinear optimal path-following
control scheme, an ADP-based path-following controller is considered
for a kinematic unicycle. The path-following technique in this paper
generates a virtual target that is then tracked by the vehicle. The
progression of the virtual target along the given path is described
by a predefined state-dependent ordinary differential equation motivated
by \cite{Lapierre.Soetanto.ea2003}. The definition of the virtual
target progression in \cite{Lapierre.Soetanto.ea2003} is desired
because it relieves a limitation on the initial condition of the vehicle
seen in \cite{Micaelli.Samson1992,Micaelli.Samson1993,Encarnacao.Pascoal2000}. 

For an infinite horizon control problem, the state associated with
the virtual target progression is unbounded, which presents several
challenges. According to the universal function approximation theorem,
a NN is a universal approximator for continuous functions on a compact
domain. Since the value function and optimal policy depend on the
unbounded path parameter, the domain of the approximation is not compact;
hence, to approximate the value function using a NN, an alternate
description of the virtual target progression that results in a compact
domain for the associated state needs to be developed. In addition,
the vehicle requires constant control effort to remain on the path;
therefore, any control policy that results in path-following also
results in infinite cost, rendering the associated control problem
ill-defined.

In this result, the progression of the virtual target is redefined
to remain on a compact domain, and the kinematics of the unicycle
are expressed as an autonomous dynamical system in terms of the error
between the vehicle and virtual target. A modified control input is
developed as the difference between the designed control and the nominal
control required to keep the vehicle on the path. The cost function
is formulated in terms of the modified control and is defined to exclude
the position of the virtual target, so that the vehicle is not penalized
for progress along the path. The resulting optimal control problem
admits admissible solutions, and an autonomous value function that
can be approximated on a compact domain, facilitating the development
of an online approximation to the optimal controller using the ADP
framework. A Lyapunov-based stability analysis is presented to establish
uniformly ultimately bounded (UUB) convergence of the vehicle to the
path while maintaining the desired speed profile and UUB convergence
of the approximate policy to the optimal policy. Simulation results
compare the policy obtained using the developed technique to an offline
numerical optimal solution for an assessment of performance.

\section{Vehicle Model}

\begin{figure}
\begin{centering}
\includegraphics[width=3.25in]{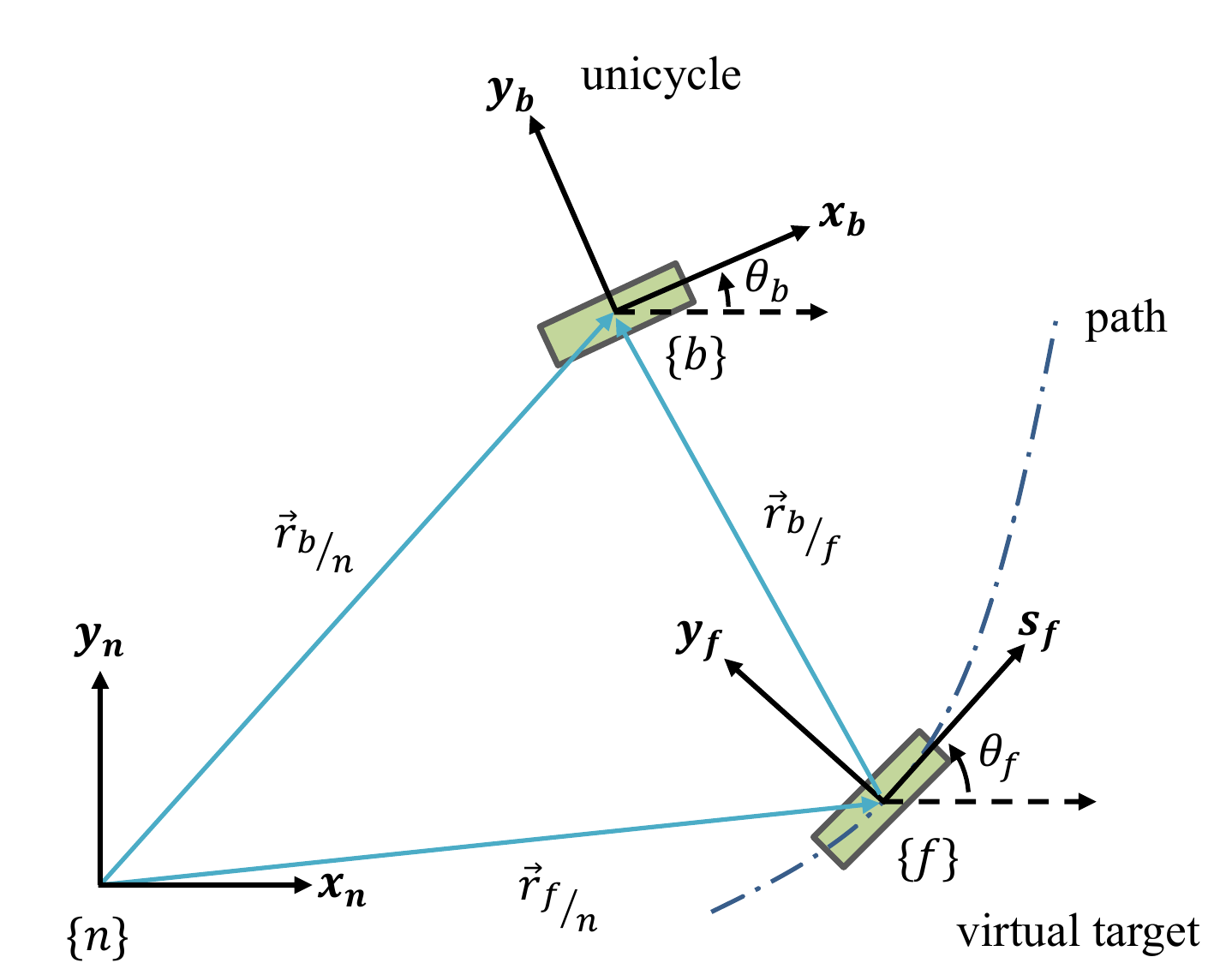}
\par\end{centering}

\caption{\label{fig:ref_frames}Description of reference frames.}
\end{figure}

Consider the nonholonomic kinematics for a unicycle given by
\begin{eqnarray}
\dot{x} & = & v\cos\theta_{b},\label{eq:basic_dyn}\\
\dot{y} & = & v\sin\theta_{b},\nonumber \\
\dot{\theta}_{b} & = & w,\nonumber 
\end{eqnarray}
where $x,y\in\mathbb{R}$ represent the vehicle's position in a plane,
and $\theta_{b}\in\mathbb{R}$ represents the angle between the vehicle's
velocity vector and the $x$-axis of the inertial frame $\left\{ n\right\} $.
The vehicle control inputs are the linear velocity $v\in\mathbb{R}$
and the angular velocity $w\in\mathbb{R}$. A time-varying body reference
frame $\left\{ b\right\} $ is defined as attached to the vehicle
with the $x$-axis aligned with the vehicle's velocity vector, as
illustrated in Figure \ref{fig:ref_frames}.

Unlike the traditional tracking problem, the path-following objective
is to move along a desired path at a specified speed profile. As typical
for this class of problems, it is convenient to express the vehicle's
kinematics in a time-varying Serret-Frenet reference frame $\left\{ f\right\} $
where the origin is fixed to a virtual target on the desired path
(cf. \cite{Lapierre.Soetanto.ea2003,Fossen2011}). Consider the following
vector equation from Figure \ref{fig:ref_frames}, $\vec{r}_{\nicefrac{b}{n}}=\vec{r}_{\nicefrac{f}{n}}+\vec{r}_{\nicefrac{b}{f}}$.
The time derivative of $\vec{r}_{\nicefrac{b}{f}}$ is $\fourIdx{n}{}{}{}{\frac{d}{dt}}\vec{r}_{\nicefrac{b}{f}}=\fourIdx{b}{}{}{}{\frac{d}{dt}}\vec{r}_{\nicefrac{b}{f}}+\vec{\omega}_{\nicefrac{b}{n}}\times\vec{r}_{\nicefrac{b}{f}}$
such that
\begin{equation}
\vec{v}_{\nicefrac{b}{n}}=\vec{v}_{\nicefrac{f}{n}}+\left(\fourIdx{b}{}{}{}{\frac{d}{dt}}\vec{r}_{\nicefrac{b}{f}}+\vec{\omega}_{\nicefrac{b}{n}}\times\vec{r}_{\nicefrac{b}{f}}\right).\label{eq:vel_equ}
\end{equation}
The velocity vector in $\left(\ref{eq:vel_equ}\right)$ can be expressed
in $\left\{ f\right\} $ as
\begin{equation}
\vec{v}_{\nicefrac{b}{n}}^{f}=\vec{v}_{\nicefrac{f}{n}}^{f}+\left(\fourIdx{b}{}{}{}{\frac{d}{dt}}\vec{r}_{\nicefrac{b}{f}}^{f}+\vec{\omega}_{\nicefrac{b}{n}}^{f}\times\vec{r}_{\nicefrac{b}{f}}^{f}\right).\label{eq:dyn_vel_eqn}
\end{equation}
The relative position of the vehicle with respect to the virtual target
is given as $\vec{r}_{\nicefrac{b}{f}}^{f}=\left[\begin{array}{ccc}
s & y & 0\end{array}\right]^{T}$ where $s\in\mathbb{R}$ is the component along the path's tangent
unit vector $s_{f}$ and $y\in\mathbb{R}$ is the component along
the path's normal unit vector $y_{f}$ taken at the virtual target's
location. From Figure \ref{fig:ref_frames}, the velocity of the vehicle
expressed in $\left\{ f\right\} $ is given as
\begin{equation}
\vec{v}_{\nicefrac{b}{n}}^{f}=R_{\theta}\left[\begin{array}{c}
v\\
0\\
0
\end{array}\right],\label{eq:dyn_vel_veh}
\end{equation}
where the rotation matrix $R_{\theta}:\mathbb{R}\rightarrow\mathbb{R}^{3\times3}$
is defined as
\[
R_{\theta}\triangleq\left[\begin{array}{ccc}
\cos\theta & -\sin\theta & 0\\
\sin\theta & \cos\theta & 0\\
0 & 0 & 1
\end{array}\right],
\]
where the angle between $\left\{ f\right\} $ and $\left\{ b\right\} $
is $\theta\triangleq\theta_{b}-\theta_{f}$ where $\theta_{f}\in\mathbb{R}$
is the angle between $s_{f}$ and $x_{n}$. Substituting $\left(\ref{eq:dyn_vel_veh}\right)$
into $\left(\ref{eq:dyn_vel_eqn}\right)$ yields
\begin{equation}
R_{\theta}\left[\begin{array}{c}
v\\
0\\
0
\end{array}\right]\!\!=\!\!\left[\begin{array}{c}
\dot{s}_{p}\\
0\\
0
\end{array}\right]\!\!+\!\!\left(\left[\begin{array}{c}
\dot{s}\\
\dot{y}\\
0
\end{array}\right]\!\!+\!\!\left[\begin{array}{c}
0\\
0\\
\kappa\left(s_{p}\right)\dot{s}_{p}
\end{array}\right]\!\!\times\!\!\left[\begin{array}{c}
s\\
y\\
0
\end{array}\right]\right),\label{eq:vector_form}
\end{equation}
where $s_{p}\in\mathbb{R}$ is the arc length the virtual target has
moved along the path, and $\kappa:\mathbb{R}\rightarrow\mathbb{R}$
is the path curvature. Based on $\left(\ref{eq:vector_form}\right)$
\begin{eqnarray}
v\cos\theta & = & \dot{s}+\left(1-\kappa y\right)\dot{s}_{p},\label{eq:dyn_raw_position}\\
v\sin\theta & = & \dot{y}+\kappa s\dot{s}_{p}.\nonumber 
\end{eqnarray}
The time derivative of the angle $\theta$ is
\begin{equation}
\fourIdx{n}{}{}{}{\frac{d}{dt}}\theta=\fourIdx{n}{}{}{}{\frac{d}{dt}}\theta_{b}-\fourIdx{n}{}{}{}{\frac{d}{dt}}\theta_{f},\label{eq:dyn_theta_dot}
\end{equation}
where $\fourIdx{n}{}{}{}{\frac{d}{dt}}\theta_{b}=w$ from $\left(\ref{eq:basic_dyn}\right)$
and $\fourIdx{n}{}{}{}{\frac{d}{dt}}\theta_{f}=\kappa\dot{s}_{p}$.
Rearranging $\left(\ref{eq:dyn_raw_position}\right)$ and augmenting
$\left(\ref{eq:dyn_theta_dot}\right)$ results in the vehicle's kinematics
expressed in $\left\{ f\right\} $ given by \cite{Lapierre.Soetanto.ea2003}

\begin{eqnarray}
\dot{s} & = & -\dot{s}_{p}\left(1-\kappa y\right)+v\cos\theta,\label{eq:fs_dyn}\\
\dot{y} & = & -\kappa\dot{s}_{p}s+v\sin\theta,\nonumber \\
\dot{\theta} & = & w-\kappa\dot{s}_{p}.\nonumber 
\end{eqnarray}

The location of the virtual target can be determined by projecting
the location of the vehicle onto the path. Assuming the projection
is well defined, $s=0$ and $\dot{s}=0$, and hence from $\left(\ref{eq:fs_dyn}\right)$
\[
\dot{s}_{p}=\frac{v\cos\theta}{1-\kappa y}.
\]
When $y=\nicefrac{1}{\kappa}$, the virtual target's velocity $\dot{s}_{p}$
is undefined. Therefore the vehicle\textquoteright{}s initial condition
is limited to a tube defined by $\nicefrac{1}{\kappa}$ along the
path, which can be restrictive for paths with large curvature values.
Motivated by the development in \cite{Lapierre.Soetanto.ea2003},
instead of projecting the location of the vehicle onto the path, the
location of the virtual target is determined by 
\begin{equation}
\dot{s}_{p}\triangleq v_{des}\cos\theta+k_{1}s,\label{eq:path_vel}
\end{equation}
where $v_{des}\in\mathbb{R}$ is a desired positive, bounded and time-invariant
speed profile, and $k_{1}\in\mathbb{R}$ is an adjustable positive
gain. This definition of the virtual target's speed eliminates the
singularity at $y=\nicefrac{1}{\kappa}$.

To facilitate the subsequent control development, we define an auxiliary
function $\phi:\mathbb{R}\rightarrow\left(-1,1\right)$ as 
\begin{equation}
\phi\triangleq\tanh\left(k_{2}s_{p}\right),\label{eq:aux_fun}
\end{equation}
 where $k_{2}\in\mathbb{R}$ is a positive gain. From $\left(\ref{eq:path_vel}\right)$
and $\left(\ref{eq:aux_fun}\right)$, the time derivative of $\phi$
is
\begin{equation}
\frac{d\phi}{ds_{p}}\frac{ds_{p}}{dt}=k_{2}\mathrm{sech}^{2}\left(\tanh^{-1}\left(\phi\right)\right)\left(v_{des}\cos\theta+k_{1}s\right).\label{eq:path_parameter}
\end{equation}
Note that the path curvature and desired speed profile can be written
as a function of $\phi$.

Based on $\left(\ref{eq:fs_dyn}\right)$ and $\left(\ref{eq:path_parameter}\right)$,
auxiliary control inputs $v_{e},w_{e}\in\mathbb{R}$ are designed
as 
\begin{eqnarray}
v_{e} & \triangleq & v-v_{ss},\label{eq:des_con}\\
w_{e} & \triangleq & w-w_{ss},\nonumber 
\end{eqnarray}
where $w_{ss}\triangleq\kappa v_{des}$ and $v_{ss}\triangleq v_{des}$
based on the control input required to remain on the path.

Substituting $\left(\ref{eq:path_vel}\right)$ and $\left(\ref{eq:des_con}\right)$
into $\left(\ref{eq:fs_dyn}\right)$, and augmenting the system state
with $\left(\ref{eq:path_parameter}\right)$, the closed-loop system
expressed in $\left\{ f\right\} $ is%
{} 
\begin{eqnarray}
\dot{s} & = & \kappa yv_{des}\cos\theta+k_{1}\kappa sy-k_{1}s+v_{e}\cos\theta\label{eq:uni_dyn}\\
\dot{y} & = & v_{des}\sin\theta-\kappa sv_{des}\cos\theta-k_{1}\kappa s^{2}+v_{e}\sin\theta\nonumber \\
\dot{\theta} & = & \kappa v_{des}-\kappa\left(v_{des}\cos\theta+k_{1}s\right)+w_{e}\nonumber \\
\dot{\phi} & = & k_{2}\mathrm{sech}^{2}\left(\tanh^{-1}\left(\phi\right)\right)\left(v_{des}\cos\theta+k_{1}s\right).\nonumber 
\end{eqnarray}
The closed-loop system in $\left(\ref{eq:uni_dyn}\right)$ can be
rewritten in the following control affine form
\begin{equation}
\dot{\zeta}=f\left(\zeta\right)+g\left(\zeta\right)u,\label{eq:dyn_gen}
\end{equation}
where $\zeta=\left[\begin{array}{cccc}
s & y & \theta & \phi\end{array}\right]^{T}\in\mathbb{R}^{4}$ is the state vector, $u=\left[\begin{array}{cc}
v_{e} & w_{e}\end{array}\right]^{T}\in\mathbb{R}^{2}$ is the control vector, and the locally Lipschitz functions $f:\mathbb{R}^{4}\rightarrow\mathbb{R}^{4}$
and $g:\mathbb{R}^{4}\rightarrow\mathbb{R}^{4\times2}$ are defined
as
\[
f\left(\zeta\right)\triangleq\left[\begin{array}{c}
\kappa yv_{des}\cos\theta+k_{1}\kappa sy-k_{1}s\\
v_{des}\sin\theta-\kappa sv_{des}\cos\theta-k_{1}\kappa s^{2}\\
\kappa v_{des}-\kappa\left(v_{des}\cos\theta+k_{1}s\right)\\
k_{2}\mathrm{sech}^{2}\left(\tanh^{-1}\left(\phi\right)\right)\left(v_{des}\cos\theta+k_{1}s\right)
\end{array}\right],
\]
\[
g\left(\zeta\right)\triangleq\left[\begin{array}{cc}
\cos\left(\theta\right) & 0\\
\sin\left(\theta\right) & 0\\
0 & 1\\
0 & 0
\end{array}\right].
\]

To facilitate the subsequent stability analysis, define a subset of
the state as $e=\left[\begin{array}{ccc}
s & y & \theta\end{array}\right]^{T}\in\mathbb{R}^{3}$.

\section{Formulation of optimal control problem}

The cost functional for the optimal control problem is defined as
\begin{equation}
J\left(\zeta,u\right)=\intop_{t}^{\infty}r\left(\zeta\left(\tau\right),u\left(\tau\right)\right)d\tau,\label{eq:cost}
\end{equation}
where $r:\mathbb{R}^{4}\rightarrow\left[0,\infty\right)$ is the local
cost defined as
\[
r\left(\zeta,u\right)=\zeta^{T}Q\zeta+u{}^{T}Ru.
\]
In $\left(\ref{eq:cost}\right)$, $R\in\mathbb{R}^{2\times2}$ is
a symmetric positive definite matrix, and $\bar{Q}\in\mathbb{R}^{4\times4}$
is defined as 
\[
\bar{Q}\triangleq\left[\begin{array}{cc}
Q & 0_{3\times1}\\
0_{1\times3} & 0
\end{array}\right],
\]
where $Q\in\mathbb{R}^{3\times3}$ is a positive definite matrix where
$\underline{q}\left\Vert \xi_{q}\right\Vert ^{2}\leq\xi_{q}^{T}Q\xi_{q}\leq\overline{q}\left\Vert \xi_{q}\right\Vert ^{2},\forall\xi_{q}\in\mathbb{R}^{3}$
where $\underline{q}$ and $\overline{q}$ are positive constants.
The infinite-time scalar value functional $V:\left[0,\infty\right)\rightarrow\left[0,\infty\right)$
is written as
\begin{equation}
V=\min_{u\in\mathcal{U}}\intop_{t}^{\infty}r\left(\zeta\left(\tau\right),u\left(\tau\right)\right)d\tau,\label{eq:value_function}
\end{equation}
where $\mathcal{U}$ is the set of admissible control policies.

The objective of the optimal control problem is to determine the optimal
policy $u^{*}$ that minimizes the cost functional $\left(\ref{eq:cost}\right)$
subject to the constraints in $\left(\ref{eq:dyn_gen}\right)$. Assuming
a minimizing policy exists and the value function is continuously
differentiable, the Hamiltonian is defined as
\begin{equation}
H\triangleq r\left(\zeta,u^{*}\right)+\frac{\partial V}{\partial\zeta}\left(f+gu^{*}\right).\label{eq:ham}
\end{equation}
The Hamilton-Jacobi-Bellman (HJB) equation is given as \cite{Kirk2004}
\begin{equation}
0=\frac{\partial V}{\partial t}+H,\label{eq:HJB}
\end{equation}
where $\frac{\partial V}{\partial t}\equiv0$ since there exists no
explicit dependence on time. The optimal policy is derived from $\left(\ref{eq:HJB}\right)$
as
\begin{equation}
u^{*}=-\frac{1}{2}R^{-1}g^{T}\left(\frac{\partial V}{\partial\zeta}\right)^{T}.\label{eq:opt_cont}
\end{equation}

The analytical expression for the optimal controller in $\left(\ref{eq:opt_cont}\right)$
requires knowledge of the value function which is the solution to
the HJB. Given the kinematics in $\left(\ref{eq:uni_dyn}\right)$
and $\left(\ref{eq:dyn_gen}\right)$, it is unclear how to determine
an analytical solution to $\left(\ref{eq:HJB}\right)$, as is generally
the case since $\left(\ref{eq:HJB}\right)$ is a nonlinear partial
differential equation; hence, the subsequent development focuses on
the development of an approximate solution.

\section{Approximate Solution}

The subsequent development is based on a neural network (NN) approximation
of the value function and optimal policy, and follows a similar structure
to \cite{Kamalapurkar.Walters.ea2013}. The development is included
here for completeness. To facilitate the use of a neural network,
a temporary assumption is made that the system state lies on a compact
set where $\zeta\left(t\right)\in\chi\subset\mathbb{R}^{4},\:\forall t\in\left[0,\infty\right)$.
This assumption is relieved in the subsequent stability analysis in
Remark \ref{If-,-then:semi-globe}, and is common in NN literature
(cf. \cite{Hornik1990,Lewis2002}).
\begin{assumption}
\label{thm:uni_fun_approx}The value function $V:\mathbb{R}^{4}\rightarrow\left[0,\infty\right)$
can be represented by a single-layer NN with $L$ neurons as
\begin{equation}
V\left(\zeta\right)=W^{T}\sigma\left(\zeta\right)+\epsilon\left(\zeta\right),\label{eq:value_nn}
\end{equation}
where $W\in\mathbb{R}^{L}$ is the ideal weight vector bounded above
by a known positive constant, $\sigma:\mathbb{R}^{4}\rightarrow\mathbb{R}^{L}$
is a bounded, continuously differentiable activation function, and
$\epsilon:\mathbb{R}^{4}\rightarrow\mathbb{R}$ is the bounded, continuously
differentiable function reconstruction error.
\end{assumption}

From $\left(\ref{eq:opt_cont}\right)$ and $\left(\ref{eq:value_nn}\right)$,
the optimal policy can be represented as 
\begin{equation}
u^{*}=-\frac{1}{2}R^{-1}g^{T}\left(\sigma'^{T}W+\epsilon'^{T}\right).\label{eq:opt_con_nn}
\end{equation}
Based on $\left(\ref{eq:value_nn}\right)$ and $\left(\ref{eq:opt_con_nn}\right)$,
the value function and optimal policy NN approximations are defined
as 
\begin{equation}
\hat{V}=\hat{W}_{c}^{T}\sigma,\label{eq:value_approx}
\end{equation}
\begin{equation}
\hat{u}=-\frac{1}{2}R^{-1}g^{T}\sigma'^{T}\hat{W}_{a},\label{eq:opt_con_approx}
\end{equation}
where $\hat{W}_{c},\:\hat{W}_{a}\in\mathbb{R}^{L}$ are estimates
of the ideal weight vector $W$. The weight estimation errors are
defined as $\tilde{W}_{c}\triangleq W-W_{c}$ and $\tilde{W}_{a}\triangleq W-W_{a}$.
The NN approximation of the Hamiltonian is given as 
\begin{equation}
\hat{H}=r\left(\zeta,\hat{u}\right)+\frac{\partial\hat{V}}{\partial\zeta}\left(f+g\hat{u}\right)\label{eq:ham_approx}
\end{equation}
by substituting $\left(\ref{eq:value_approx}\right)$ and $\left(\ref{eq:opt_con_approx}\right)$
into $\left(\ref{eq:ham}\right)$. The Bellman error $\delta\in\mathbb{R}$
is defined as the error between the optimal and approximate Hamiltonian
and is given as
\begin{equation}
\delta\triangleq\hat{H}-H,\label{eq:bell_err}
\end{equation}
where $H\equiv0$. Therefore, the Bellman error can be written in
a measurable form as 
\[
\delta=r\left(\zeta,\hat{u}\right)+\hat{W_{c}}^{T}\omega,
\]
where $\omega\triangleq\sigma'\left(f+g\hat{u}\right)\in\mathbb{R}^{L}$.
\begin{assumption}
\label{thm:concurrent}There exists a set of sampled data points $\left\{ \zeta_{j}\in\chi|j=1,2,\ldots,N\right\} $
such that $\forall t\in\left[0,\infty\right)$,
\begin{equation}
\mathrm{rank}\left(\sum_{j=1}^{N}\frac{\omega_{j}\omega_{j}^{T}}{p_{j}}\right)=L,\label{eq:rank_cond}
\end{equation}
where $p_{j}\triangleq\sqrt{1+\omega_{j}^{T}\omega_{j}}$ denotes
the normalization constant, and $\omega_{j}$ is evaluated at the
specified data point $\zeta_{j}$. 
\end{assumption}

In general, the rank condition in $\left(\ref{eq:rank_cond}\right)$
cannot be guaranteed to hold a priori. However, heuristically, the
condition can be met by sampling redundant data, i.e., $N\gg L$.
Based on Assumption \ref{thm:concurrent}, it can be shown that $\sum_{j=1}^{N}\frac{\omega_{j}\omega_{j}^{T}}{p_{j}}>0$
such that
\[
\underline{c}\left\Vert \xi_{c}\right\Vert ^{2}\leq\xi_{c}^{T}\left(\sum_{j=1}^{n}\frac{\omega_{j}\omega_{j}^{T}}{p_{j}}\right)\xi_{c}\leq\overline{c}\left\Vert \xi_{c}\right\Vert ^{2},\:\forall\xi_{c}\in\mathbb{R}^{4}
\]
even in the absence of persistent excitation \cite{Chowdhary.Johnson2011a,Chowdhary.Yucelen.ea2012}.

The adaptive update law for $\hat{W}_{c}$ in $\left(\ref{eq:value_approx}\right)$
is given by 
\begin{equation}
\dot{\hat{W}}_{c}=-\eta_{c1}\frac{\partial\delta}{\partial\hat{W}_{c}}\frac{\delta}{p}-\frac{\eta_{c2}}{N}\sum_{j=1}^{N}\frac{\partial\delta_{j}}{\partial\hat{W}_{c}}\frac{\delta_{j}}{p_{j}},\label{eq:update_value}
\end{equation}
where $\eta_{c1},\:\eta_{c2}\in\mathbb{R}$ are positive adaptation
gains, $\frac{\partial\delta}{\partial\hat{W}_{c}}=\omega$ is the
regressor matrix, and $p\triangleq\sqrt{1+\omega^{T}\omega}$ is a
normalization constant. The policy NN update law for $\hat{W}_{a}$
in $\left(\ref{eq:opt_con_approx}\right)$ is given by
\begin{equation}
\dot{\hat{W}}_{a}=\mathrm{proj}\left\{ -\eta_{a}\left(\hat{W}_{a}-\hat{W}_{c}\right)\right\} ,\label{eq:update_policy}
\end{equation}
where $\eta_{a}\in\mathbb{R}$ is a positive gain, and $\mathrm{proj}\left\{ \cdot\right\} $
is a smooth projection operator \cite{Dixon2003}. Using Assumption
\ref{thm:uni_fun_approx} and the properties of the projection operator,
the policy NN weight estimation errors are bounded above by positive
constants.

\section{Stability Analysis}

To facilitate the subsequent stability analysis, an unmeasurable from
of the Bellman error can be written using $\left(\ref{eq:ham}\right)$,
$\left(\ref{eq:ham_approx}\right)$, and $\left(\ref{eq:bell_err}\right)$,
as%
\begin{equation}
\delta=-\tilde{W}_{c}^{T}\omega-\epsilon'f+\frac{1}{2}\epsilon'G\sigma'^{T}W+\frac{1}{4}\tilde{W}_{a}^{T}G_{\sigma}\tilde{W}_{a}+\frac{1}{4}\epsilon'G\epsilon'^{T},\label{eq:unmeas_bell}
\end{equation}
where $G\triangleq gR^{-1}g^{T}\in\mathbb{R}^{4\times4}$ and $G_{\sigma}\triangleq\sigma'G\sigma'^{T}\in\mathbb{R}^{L\times L}$
are symmetric, positive semi-definite matrices. Similarly, at the
sampled points the Bellman error can be written as
\begin{equation}
\delta_{j}=-\tilde{W}_{c}^{T}\omega_{j}+\frac{1}{4}\tilde{W}_{a}^{T}G_{\sigma j}\tilde{W}_{a}+E_{j},\label{eq:sam_unmes_bell}
\end{equation}
where $E_{j}\triangleq\frac{1}{2}\epsilon_{j}'G_{j}\sigma_{j}'^{T}W+\frac{1}{4}\epsilon_{j}'G_{j}\epsilon_{j}'^{T}-\epsilon_{j}'f_{j}\in\mathbb{R}$.

The function $f$ on the compact set $\chi$ is Lipschitz continuous,
and therefore bounded by
\[
\left\Vert f\left(\zeta\right)\right\Vert \leq L_{f}\left\Vert \zeta\right\Vert ,\:\forall\zeta\in\chi,
\]
where $L_{f}$ is the positive Lipschitz constant, and the normalized
regressor in $\left(\ref{eq:update_value}\right)$ is upper bounded
by $\left\Vert \frac{\omega}{p}\right\Vert \leq1$.

The augmented equations of motion in $\left(\ref{eq:uni_dyn}\right)$
present a unique challenge with respect to the value function $V$
which is utilized as a Lyapunov function in the stability analysis.
To prevent penalizing the vehicle progression along the path, the
path parameter $\phi$ is removed from the cost function with the
introduction of a positive semi-definite state weighting matrix $\bar{Q}$.
However, since $\bar{Q}$ is positive semi-definite, efforts are required
to ensure the value function is positive definite. To address this
challenge, the fact that the value function can be interpreted as
a time-invariant map $V:\mathbb{R}^{4}\rightarrow\left[0,\infty\right)$
or a time-varying map $V:\mathbb{R}^{3}\times\left[0,\infty\right)\rightarrow\left[0,\infty\right)$
is exploited. Specifically, the time-invariant map facilitates the
optimal policy development while the time-varying map facilitates
the stability analysis. Lemma \ref{lem:pos_def} is used to show that
the time-varying map is a positive definite and decrescent function
for use as a Lyapunov function. 
\begin{lem}
\label{lem:pos_def}Let $B_{a}$ denote a closed ball around the origin
with the radius $a\in\left[0,\infty\right)$, The optimal value function
$V:\mathbb{R}^{3}\times\left[0,\infty\right)\rightarrow\mathbb{R}$
satisfies the following properties
\[
V\left(0,t\right)=0,
\]
\[
\underline{\upsilon}\left(\left\Vert e\right\Vert \right)\leq V\left(e,t\right)\leq\overline{\upsilon}\left(\left\Vert e\right\Vert \right),
\]
$\forall t\in\left[0,\infty\right)$ and $\forall e\in B_{a}\subset\chi$
where $\underline{\upsilon}:\left[0,a\right]\rightarrow\left[0,\infty\right)$
and $\overline{\upsilon}:\left[0,a\right]\rightarrow\left[0,\infty\right)$
are class $\mathcal{K}$ functions.\end{lem}
\begin{IEEEproof}
The proof follows similarly to the proof of Lemma 2 in \cite{arxivKamalapurkar.Dinh.ea2013a}.
\end{IEEEproof}

\begin{thm}
If Assumptions \ref{thm:uni_fun_approx} and \ref{thm:concurrent}
hold, and the following sufficient conditions are satisfied
\begin{equation}
\underline{q}>\frac{\eta_{c1}\overline{\left\Vert \epsilon'\right\Vert }L_{f}}{2},\label{eq:gain_cond_q}
\end{equation}
\begin{equation}
\underline{c}>\frac{N\eta_{a}}{2\eta_{c2}}+\frac{N\eta_{c1}\overline{\left\Vert \epsilon'\right\Vert }L_{f}}{2\eta_{c2}},\label{eq:gain_con_c}
\end{equation}
 where $\overline{\left\Vert \cdot\right\Vert }\triangleq\sup_{\zeta}\left\Vert \cdot\right\Vert $
and $Z\triangleq\left[\begin{array}{ccc}
e^{T} & \tilde{W}_{c}^{T} & \tilde{W}_{a}^{T}\end{array}\right]^{T}\in\mathcal{Z}\subset\mathbb{R}^{3}\times\mathbb{R}^{L}\times\mathbb{R}^{L}$, then the policy in $\left(\ref{eq:opt_con_approx}\right)$ with
the NN update laws in $\left(\ref{eq:update_value}\right)$ and $\left(\ref{eq:update_policy}\right)$
guarantee UUB regulation of vehicle to the virtual target and UUB
convergence of the approximate policy to the optimal policy.\end{thm}
\begin{IEEEproof}
Consider the continuously differentiable, positive definite candidate
Lyapunov function $V_{L}:\mathbb{R}^{3}\times\mathbb{R}^{L}\times\mathbb{R}^{L}\rightarrow\left[0,\infty\right)$
given as
\[
V_{L}\left(Z,t\right)=V\left(e,t\right)+\frac{1}{2}\tilde{W}_{c}^{T}\tilde{W}_{c}+\frac{1}{2}\tilde{W}_{a}^{T}\tilde{W}_{a}.
\]
Using Lemma \ref{lem:pos_def}, the candidate Lyapunov function can
be bounded by
\begin{equation}
\underline{\upsilon_{L}}\left(\left\Vert Z\right\Vert \right)\leq V_{L}\leq\overline{\upsilon_{L}}\left(\left\Vert Z\right\Vert \right),\:\forall Z\in B_{b},\:\forall t\in\left[0,\infty\right),\label{eq:lyp_can}
\end{equation}
 where $\underline{\upsilon_{L}},\overline{\upsilon_{L}}:\left[0,b\right]\rightarrow\left[0,\infty\right)$
are class $\mathcal{K}$ functions and $B_{b}\subset\mathcal{Z}$
denotes a ball of radius $b\in\left[0,\infty\right)$ around the origin.

The time derivative of the candidate Lyapunov function is%
\begin{comment}
\[
V_{L}=V+\frac{1}{2}\tilde{W}_{c}^{T}\tilde{W}_{c}+\frac{1}{2}\tilde{W}_{a}^{T}\tilde{W}_{a}
\]
\[
\dot{V}_{L}=\frac{d}{dt}V+\frac{1}{2}\frac{d}{dt}\left(\tilde{W}_{c}^{T}\tilde{W}_{c}\right)+\frac{1}{2}\frac{d}{dt}\left(\tilde{W}_{a}^{T}\tilde{W}_{a}\right)
\]
\[
\dot{V}_{L}=\frac{\partial V}{\partial\zeta}\frac{d\zeta}{dt}-\tilde{W}_{c}^{T}\dot{\hat{W}}_{c}-\tilde{W}_{a}^{T}\dot{\hat{W}}_{a}
\]
\[
\dot{V}_{L}=\frac{\partial V}{\partial\zeta}\left(f+g\hat{u}\right)-\tilde{W}_{c}^{T}\dot{\hat{W}}_{c}-\tilde{W}_{a}^{T}\dot{\hat{W}}_{a}
\]
\end{comment}
{} 
\[
\dot{V}_{L}=\frac{\partial V}{\partial\zeta}f+\frac{\partial V}{\partial\zeta}g\hat{u}-\tilde{W}_{c}^{T}\dot{\hat{W}}_{c}-\tilde{W}_{a}^{T}\dot{\hat{W}}_{a}.
\]
Using $\left(\ref{eq:HJB}\right)$, $\frac{\partial V}{\partial\zeta}f=-\frac{\partial V}{\partial\zeta}gu^{*}-r\left(\zeta,u^{*}\right)$.
Then, 
\[
\dot{V}_{L}=\frac{\partial V}{\partial\zeta}g\hat{u}-\frac{\partial V}{\partial\zeta}gu^{*}-r\left(\zeta,u^{*}\right)-\tilde{W}_{c}^{T}\dot{\hat{W}}_{c}-\tilde{W}_{a}^{T}\dot{\hat{W}}_{a}.
\]
 Substituting for $\left(\ref{eq:update_value}\right)$ and $\left(\ref{eq:update_policy}\right)$
yields
\begin{eqnarray*}
\dot{V}_{L} & = & -e^{T}Qe-u^{*}Ru^{*}+\frac{\partial V}{\partial\zeta}g\hat{u}-\frac{\partial V}{\partial\zeta}gu^{*}\\
 &  & +\tilde{W}_{c}^{T}\left[\eta_{c1}\frac{\omega^{T}}{p}\delta+\frac{\eta_{c2}}{N}\sum_{j=1}^{N}\frac{\omega_{j}}{p_{j}}\delta_{j}\right]\\
 &  & +\tilde{W}_{a}^{T}\eta_{a}\left(\hat{W}_{a}-\hat{W}_{c}\right).
\end{eqnarray*}
 Using Young's inequality, $\left(\ref{eq:value_nn}\right)$, $\left(\ref{eq:opt_con_nn}\right)$,
$\left(\ref{eq:opt_con_approx}\right)$, $\left(\ref{eq:unmeas_bell}\right)$,
and $\left(\ref{eq:sam_unmes_bell}\right)$ the Lyapunov derivative
can be upper bounded as%
\begin{eqnarray*}
\dot{V}_{L} & \leq & -\varphi_{e}\left\Vert e\right\Vert ^{2}-\varphi_{c}\left\Vert \tilde{W}_{c}\right\Vert ^{2}-\varphi_{a}\left\Vert \tilde{W}_{a}\right\Vert ^{2}\\
 &  & +\iota_{c}\left\Vert \tilde{W}_{c}\right\Vert +\iota_{a}\left\Vert \tilde{W}_{a1}\right\Vert +\iota,
\end{eqnarray*}
where

\[
\varphi_{e}=\underline{q}-\frac{\eta_{c1}\overline{\left\Vert \epsilon'\right\Vert }L_{f}}{2},
\]
\[
\varphi_{c}=\frac{\eta_{c2}}{N}\underline{c}-\frac{\eta_{a}}{2}-\frac{\eta_{c1}\overline{\left\Vert \epsilon'\right\Vert }L_{f}}{2},
\]

\[
\varphi_{a}=\frac{\eta_{a}}{2},
\]
\begin{eqnarray*}
\iota_{c} & =\sup_{\zeta\in\chi} & \left\Vert \frac{\eta_{c2}}{4N}\sum_{j=1}^{N}\tilde{W}_{a}^{T}G_{\sigma j}\tilde{W}_{a}+\frac{\eta_{c1}}{4}\tilde{W}_{a}^{T}G_{\sigma}\tilde{W}_{a}\right.\\
 &  & +\frac{\eta_{c1}}{2}\epsilon'G\sigma'^{T}W+\frac{\eta_{c1}}{4}\epsilon'G\epsilon'^{T}\\
 &  & \left.+\frac{\eta_{c2}}{N}\sum_{j=1}^{N}E_{j}+\eta_{c1}\overline{\left\Vert \epsilon'\right\Vert }L_{f}\right\Vert ,
\end{eqnarray*}
\[
\iota_{a}=\sup_{\zeta\in\chi}\left\Vert \frac{1}{2}G_{\sigma}W+\frac{1}{2}\sigma'G\epsilon'^{T}\right\Vert ,
\]
\[
\iota=\sup_{\zeta\in\chi}\left\Vert \frac{1}{4}\epsilon'G\epsilon'^{T}\right\Vert .
\]
The constants $\varphi_{e}$, $\varphi_{c}$, and $\varphi_{a}$ are
positive if the inequalities
\[
\underline{q}>\frac{\eta_{c1}\overline{\left\Vert \epsilon'\right\Vert }L_{f}}{2},
\]
\[
\underline{c}>\frac{N\eta_{a}}{2\eta_{c2}}+\frac{N\eta_{c1}\overline{\left\Vert \epsilon'\right\Vert }L_{f}}{2\eta_{c2}}
\]
are satisfied. Completing the squares, the upper bound on the Lyapunov
derivative can be written as
\begin{eqnarray*}
\dot{V}_{L} & \leq & -\varphi_{e}\left\Vert e\right\Vert ^{2}-\frac{\varphi_{c}}{2}\left\Vert \tilde{W}_{c}\right\Vert ^{2}-\frac{\varphi_{a}}{2}\left\Vert \tilde{W}_{a}\right\Vert ^{2}\\
 &  & +\frac{\iota_{c}^{2}}{2\varphi_{c}}+\frac{\iota_{a}^{2}}{2\varphi_{a}}+\iota,
\end{eqnarray*}
which can be upper bounded as
\begin{equation}
\dot{V}_{L}\leq-\alpha\left\Vert Z\right\Vert ^{2},\forall\left\Vert Z\right\Vert \geq K>0,\label{eq:lyp_final}
\end{equation}
where $\alpha\in\mathbb{R}$ is a positive constant and 

\[
K\triangleq\sqrt{\frac{\iota_{c}^{2}}{2\alpha\varphi_{c}}+\frac{\iota_{a}^{2}}{2\alpha\varphi_{a}}+\frac{\iota}{\alpha}}.
\]
Invoking Theorem 4.18 in \cite{Khalil2002}, $Z$ is UUB. Based on
the definition of $Z$, and the inequalities in $\left(\ref{eq:lyp_can}\right)$
and $\left(\ref{eq:lyp_final}\right)$, $e,\:\tilde{W}_{c},\:\tilde{W}_{a}\in\mathcal{L}_{\infty}$.
Since $\phi\in\mathcal{L}_{\infty}$ by definition in $\left(\ref{eq:path_parameter}\right)$,
then $\zeta\in\mathcal{L}_{\infty}$. $\hat{W}_{c},\:\hat{W}_{a}\in\mathcal{L}_{\infty}$
follows from the definition of $W$. From $\left(\ref{eq:value_approx}\right)$
and $\left(\ref{eq:opt_con_approx}\right)$, $\hat{V},\:\hat{u}\in\mathcal{L}_{\infty}$.
From $\left(\ref{eq:dyn_gen}\right)$, $\dot{\zeta}\in\mathcal{L}_{\infty}$.
By the definition in $\left(\ref{eq:bell_err}\right)$, $\delta\in\mathcal{L}_{\infty}$.
From $\left(\ref{eq:update_value}\right)$ and $\left(\ref{eq:update_policy}\right)$,
$\dot{\hat{W}}_{a},\:\dot{\hat{W}}_{c}\in\mathcal{L}_{\infty}$.\end{IEEEproof}
\begin{rem}
\label{If-,-then:semi-globe}If $\left\Vert Z\left(0\right)\right\Vert \geq K$,
then $\dot{V}_{L}\left(Z\left(0\right)\right)<0.$ There exists an
$\varepsilon\in\left[0,\infty\right)$ such that $V_{L}\left(Z\left(\varepsilon\right)\right)<V_{L}\left(Z\left(0\right)\right).$
Using $\left(\ref{eq:lyp_can}\right)$, $\underline{\upsilon_{L}}\left(\left\Vert Z\left(\varepsilon\right)\right\Vert \right)\leq V_{L}\leq\overline{\upsilon_{L}}\left(\left\Vert Z\left(0\right)\right\Vert \right)$.
Rearranging terms, $\left\Vert Z\left(\varepsilon\right)\right\Vert <\underline{\upsilon_{L}}^{-1}\left(\overline{\upsilon_{L}}\left(\left\Vert Z\left(0\right)\right\Vert \right)\right).$
Hence, $Z\left(\varepsilon\right)\in\mathcal{L}_{\infty}$. It can
be shown by induction that $Z\left(t\right)\in\mathcal{L}_{\infty},\:\forall t\in\left[0,\infty\right)$
when $\left\Vert Z\left(0\right)\right\Vert >K$. Using a similar
argument when $\left\Vert Z\left(0\right)\right\Vert <K,\:\left\Vert Z\left(t\right)\right\Vert <\underline{\upsilon_{L}}^{-1}\left(\overline{\upsilon_{L}}\left(K\right)\right)$.
Therefore, $Z\left(t\right)\in\mathcal{L}_{\infty},\:\forall t\in\left[0,\infty\right)$
when $\left\Vert Z\left(0\right)\right\Vert <K$. Since $Z\left(t\right)\in\mathcal{L}_{\infty}\:\forall t\in\left[0,\infty\right)$,
and $\phi\in\mathcal{L}_{\infty}$ by definition, the state $\zeta$
lies on the compact set $\chi$ where $\chi\triangleq\left\{ \zeta\in\mathbb{R}^{4}|\left\Vert \zeta\right\Vert \leq\underline{\upsilon_{L}}^{-1}\left(\overline{\upsilon_{L}}\left(\max\left(\left\Vert Z\left(0\right)\right\Vert ,K\right)\right)\right)\right\} $.
\end{rem}

\section{Simulation}

\begin{figure}
\begin{centering}
\includegraphics[width=3.25in]{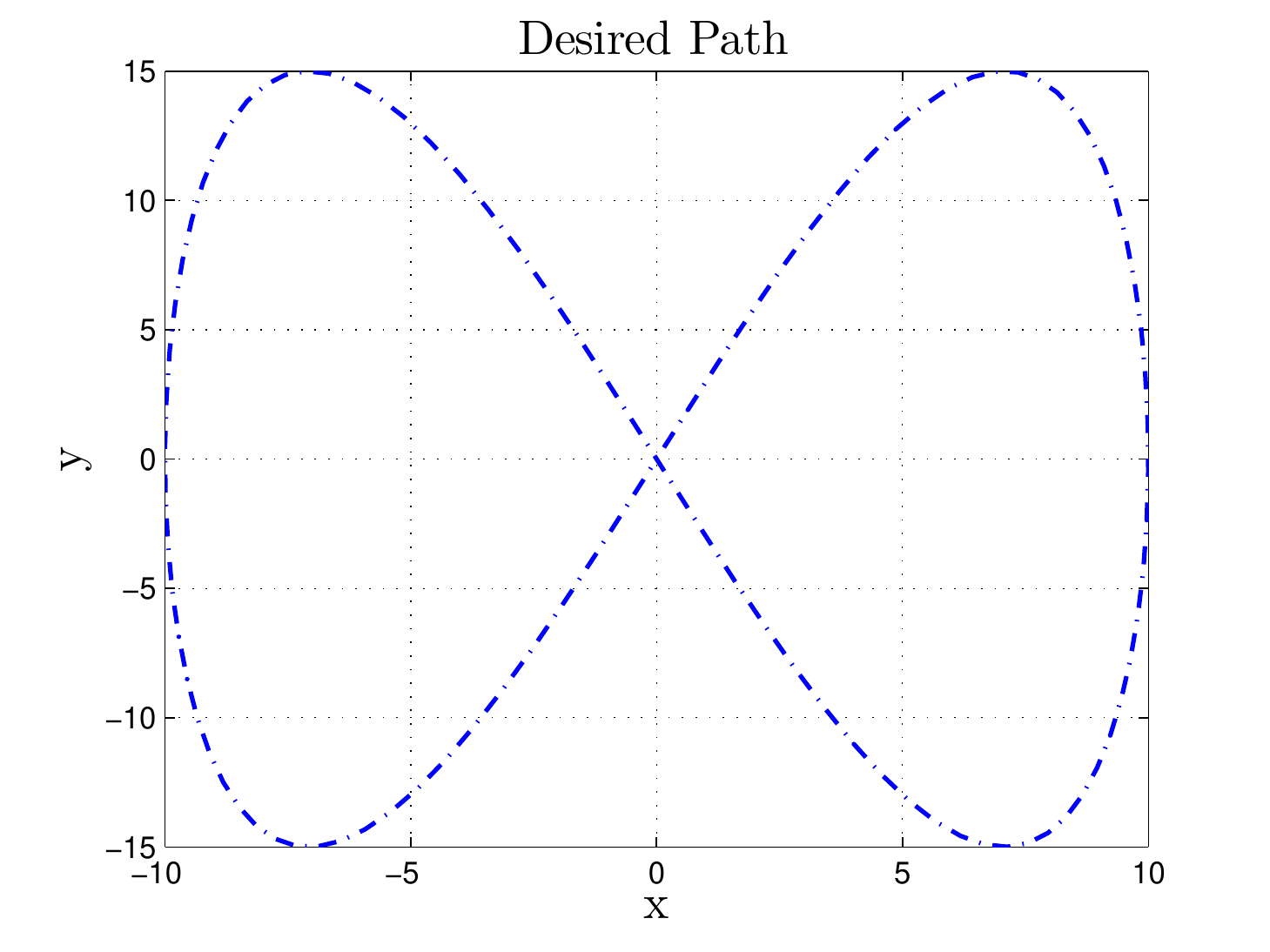}
\par\end{centering}

\caption{\label{fig:desired_path}The desired path expressed in $\left\{ n\right\} $.
From the selected initial condition, the start of the path is $x_{des}\left(0\right)=0$
and $y_{des}\left(0\right)=0$.}

\end{figure}

To demonstrate the performance of the developed ADP-based controller
a numerical simulation is performed using the kinematics in $\left(\ref{eq:basic_dyn}\right)$.
As illustrated in Figure \ref{fig:desired_path}, the equations describing
the desired path are selected as 
\begin{eqnarray}
x_{des} & = & 10\sin\left(s_{p}\right),\label{eq:des_path}\\
y_{des} & = & 15\sin\left(2s_{p}\right).\nonumber 
\end{eqnarray}
The results of the simulation are compared to a solution obtained
by the offline optimal solver GPOPS \cite{Rao.Benson.ea2010}. The
basis for the value function approximation is selected as
\[
\sigma=\left[\begin{array}{ccccccccc}
\!\!\zeta_{1}\zeta_{2} & \!\!\zeta_{1}\zeta_{3} & \!\!\zeta_{1}\zeta_{4} & \!\!\zeta_{2}\zeta_{3} & \!\!\zeta_{2}\zeta_{4} & \!\!\zeta_{3}\zeta_{4} & \!\!\zeta_{1} & \!\!\zeta_{2} & \!\!\zeta_{3}\end{array}\right].
\]
The sampled data points are selected on a $3\times3\times3\times3$
grid about the origin. The quadratic cost weighting matrices are selected
as $Q=I_{3\times3}$ and $R=I_{2\times2}$. The learning gains and
adjustable gains in the auxiliary function in $\left(\ref{eq:path_parameter}\right)$
are selected as $\eta_{c1}=0.5,\:\eta_{c1}=10,\:\eta_{a}=5.0,\: k_{1}=0.1,$
and $k_{2}=0.05$. The desired speed profile is selected as $v_{des}=0.5$.
The policy and value function NN weight estimates are initialized
to a set of stabilizing gains as $\hat{W}_{c}\left(0\right)=\hat{W}_{a}\left(0\right)=\left[\begin{array}{ccccccccc}
0 & 0 & 0 & 0.5 & 0 & 0 & 0.5 & 0 & 1.0\end{array}\right]^{T}$, and the initial condition of the unicycle is selected as $\zeta\left(0\right)=\left[\begin{array}{cccc}
-0.5 & 0.25 & -\frac{\pi}{6} & 0\end{array}\right]^{T}$.
\begin{figure}
\begin{centering}
\includegraphics[width=3.25in]{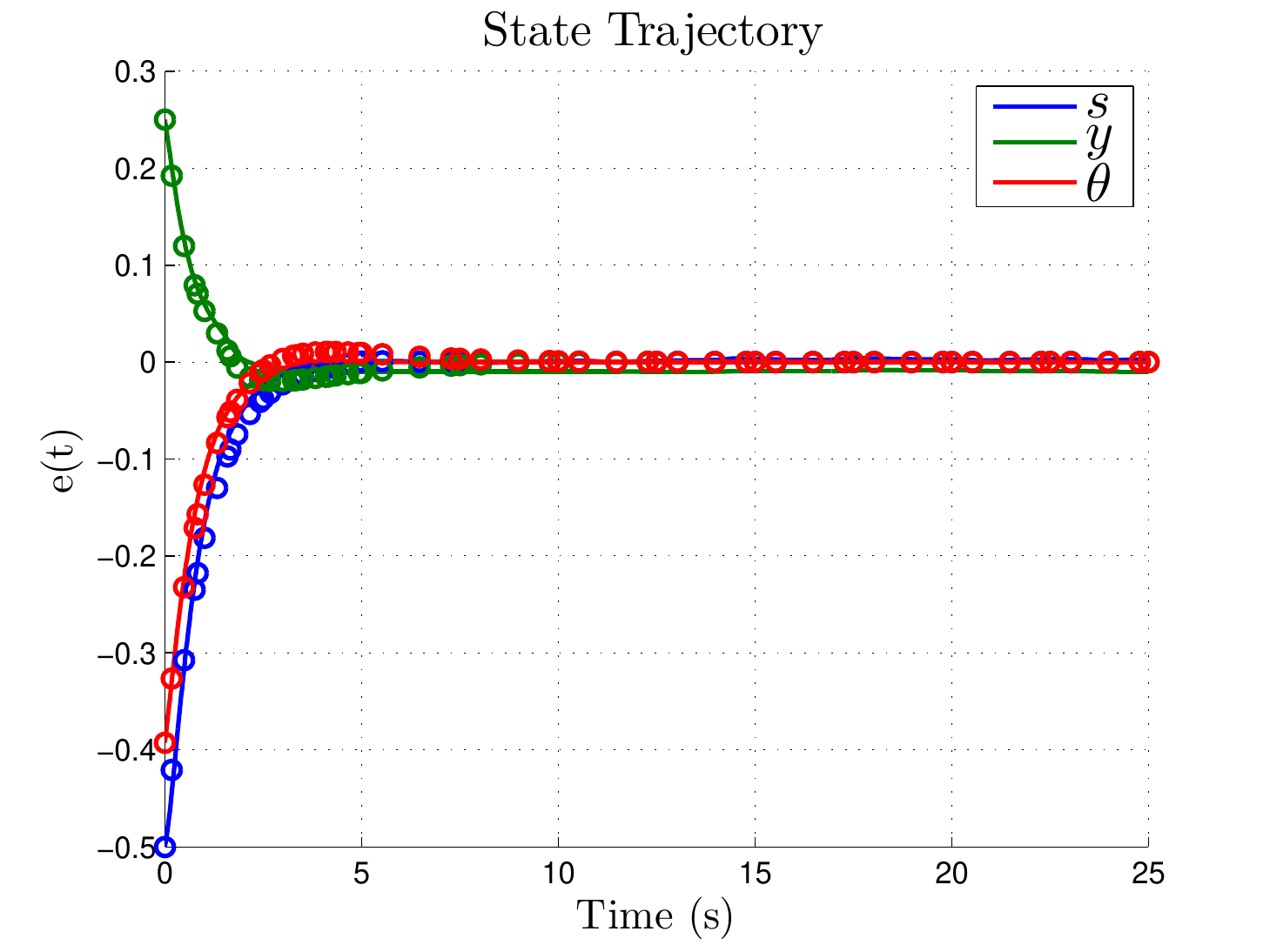}
\par\end{centering}

\caption{\label{fig:state_traj}The state trajectory generated by the developed
method is shown as solid lines, and the collocation points from GPOPS
as circular markers.}

\end{figure}
\begin{figure}
\begin{centering}
\includegraphics[width=3.25in]{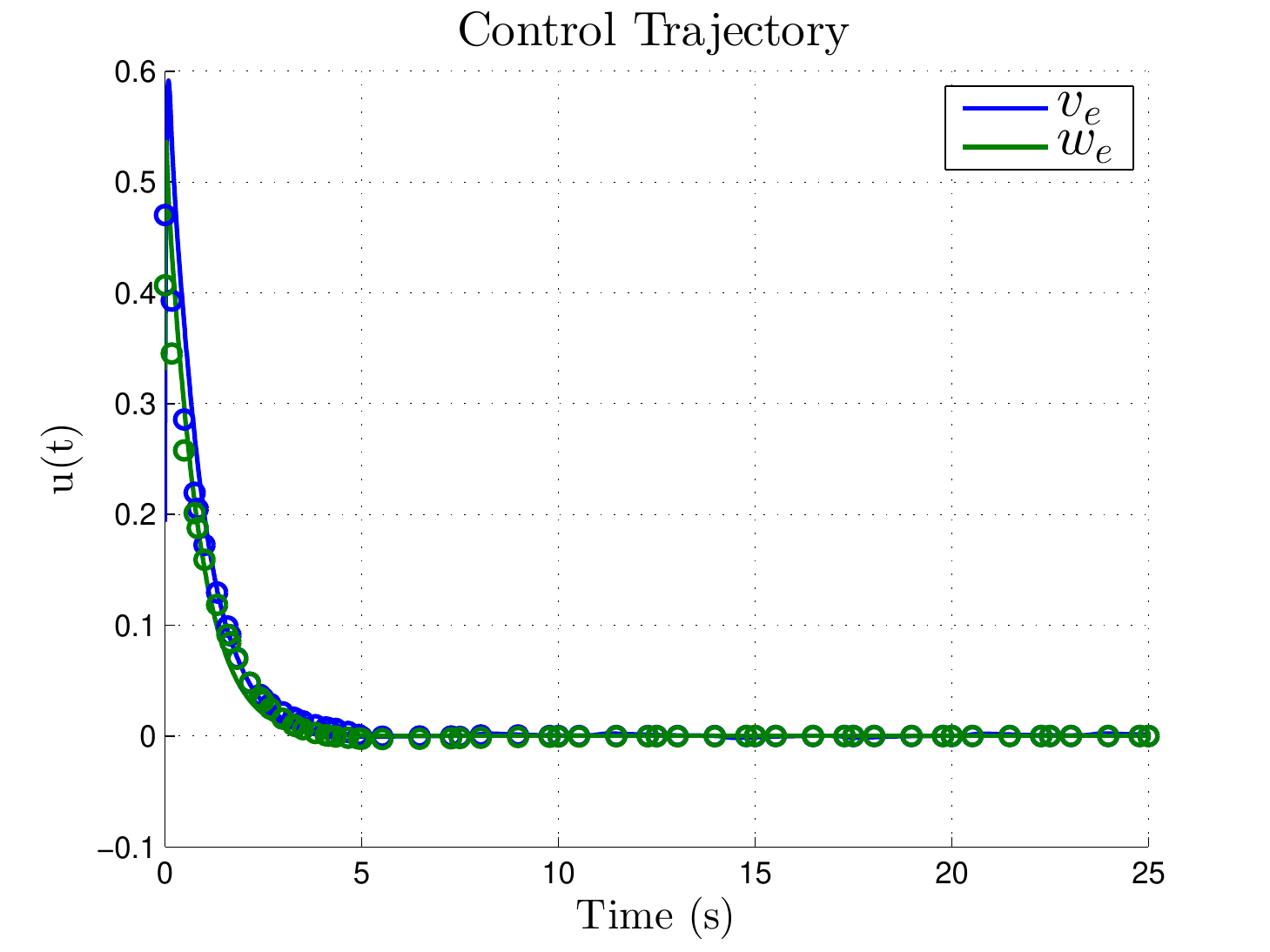}
\par\end{centering}

\caption{\label{fig:con_traj}The control trajectory generated by the developed
method is shown as solid lines, and the collocation points from GPOPS
as circular markers.}

\end{figure}
\begin{figure}
\begin{centering}
\includegraphics[width=1.625in]{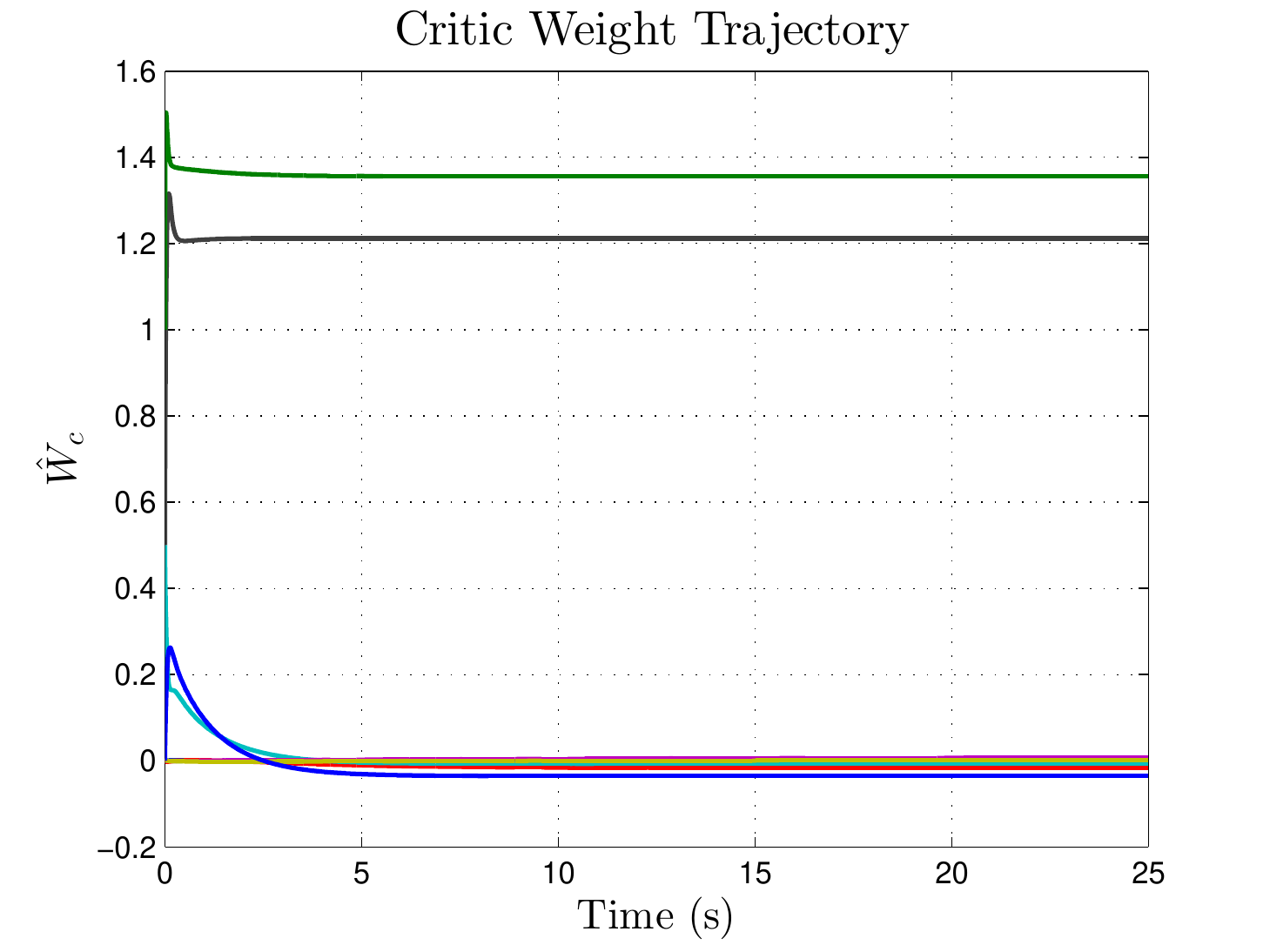}\includegraphics[width=1.625in]{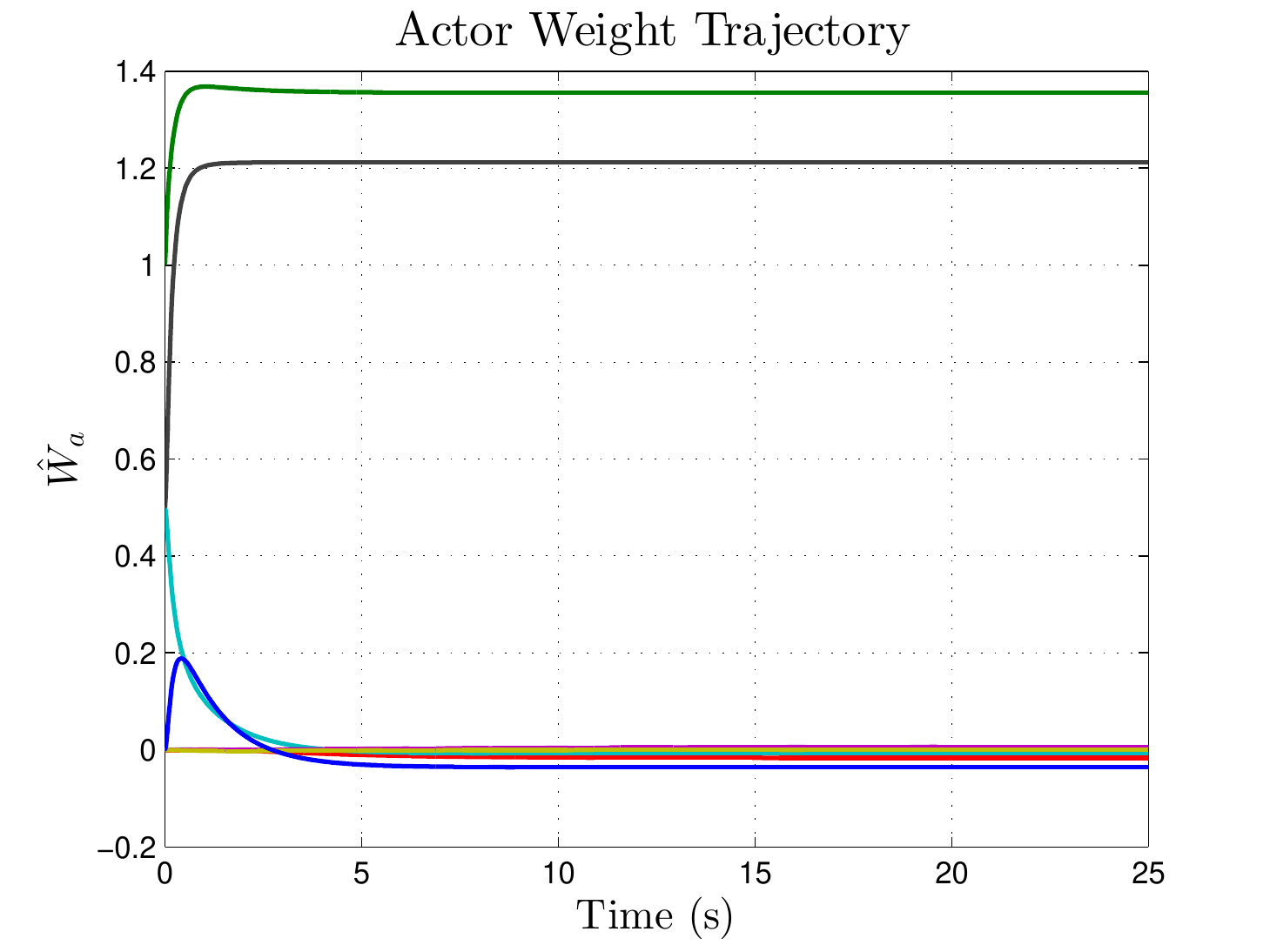}
\par\end{centering}

\caption{\label{fig:NN_traj}NN weight estimate trajectories generated by the
developed method.}
\end{figure}

Figure \ref{fig:state_traj} and \ref{fig:con_traj} show that the
state and control trajectories (denoted by solid lines) approach the
solution found using an offline optimal solver (denoted by circular
markers), and Figure \ref{fig:NN_traj} shows the NN weight estimates
converge to steady state values. Although the true values of the ideal
NN network weights are unknown, the system trajectories obtained using
the developed method correlate with the system trajectories of the
offline optimal solver.

\section{Conclusion}

An online approximation of a path-following optimal controller is
developed for a unicycle. Approximate dynamic programming is used
to approximate the solution to the HJB equation without the need for
persistence of excitation. A concurrent learning based gradient descent
adaptive update law approximates the values function. A Lyapunov-based
stability analysis proves UUB convergence of the vehicle to the desired
path while maintaining the desired speed profile, and UUB convergence
of the approximate policy to the optimal policy. Simulation results
demonstrate the utility of the proposed controller. 

A current limitation of the developed approach is that the performance
is dependent on the selection of basis functions for value function
approximation, which is evident in the UUB bound. The linear basis
functions selected for the simulation results in this paper generate
a good approximation of the value function in a local neighborhood
of the desired path. With the selection of a basis function that more
fully describes the value function, improved convergence to the desired
path throughout the state space could be achieved. Future work also
includes extension of the developed ADP-based path-following technique
to systems with additional constraints.

\bibliographystyle{IEEEtran}
\bibliography{encr,master,ncr}

\end{document}